\documentclass[twocolumn,aps,prl,groupedaddress]{revtex4}
\usepackage{graphicx}
\usepackage{color}
\usepackage{amsmath}
\usepackage{marvosym}
\usepackage{bm}

\newcommand{\be}{\begin{equation}}
\newcommand{\ee}{\end{equation}}
\newcommand{\bea}{\begin{eqnarray}}
\newcommand{\eea}{\end{eqnarray}}

\newcommand{\la}{\langle}
\newcommand{\ra}{\rangle}

\renewcommand{\vec}[1]{{\bf #1}}
\renewcommand{\epsilon}{\varepsilon}

\begin{document}

\title{Chiral plasmons without magnetic field} 
\author{Justin C. W. Song$^{1}$ and Mark S. Rudner$^2$}
\affiliation{$^1$ Walter Burke Institute of Theoretical Physics, Institute of Quantum Information and Matter, and Department of Physics, California Institute of Technology, Pasadena, CA 91125 USA}
\affiliation{$^2$ Center for Quantum Devices and Niels Bohr International Academy, Niels Bohr Institute, University of Copenhagen, 2100 Copenhagen, Denmark}

\begin{abstract}
Plasmons, the collective oscillations of interacting electrons, possess emergent properties that dramatically alter the optical response of metals. We predict the existence of a new class of plasmons  -- chiral Berry plasmons (CBPs) -- for a wide range of two-dimensional metallic systems including 
gapped Dirac materials. As we show, in these materials the interplay between Berry curvature and electron-electron interactions yields {\it chiral plasmonic modes} at zero magnetic field. The CBP modes are confined to system boundaries, even in the absence of topological edge states, with chirality manifested in split energy dispersions for oppositely directed plasmon waves. We unveil a rich CBP phenomenology and propose setups for realizing 
them, including in anomalous Hall metals and optically-pumped 2D Dirac materials. Realization of CBPs will offer a new paradigm for magnetic field-free, sub-wavelength optical non-reciprocity, in the mid IR-THz range, with tunable splittings as large as tens of THz, as well as sensitive all-optical diagnostics of topological bands. 
\end{abstract}
\maketitle

\vspace{2mm}
In electronic systems, {\it chirality} expresses the system's ability to discriminate between forward and backwards propagation of electronic signals along certain directions.
This technologically useful and hotly sought-after property 
can be achieved through the application of external magnetic fields.
However, the need for strong applied fields ``on-chip'' brings many challenges for applications.
Recently, materials exhibiting chirality in the absence of a magnetic field have
started to gain prominence. These include metals exhibiting anomalous-~\cite{nagaosa} and quantum anomalous-~\cite{haldane,yu2010,nagaosa2011,qahe,wang14} Hall effects, as well as 
non-magnetic materials pushed out-of-equilibrium, 
where for example a zero-field charge Hall effect was recently demonstrated~\cite{mak2014}. In each case, zero-field chirality arises from Bloch band Berry curvature, a fundamental property of Bloch eigenstates that dramatically affects single-particle electronic motion and material responses~\cite{dixiao,hasan,nagaosa}.

\begin{figure}[ht!]
\includegraphics[width=\columnwidth]{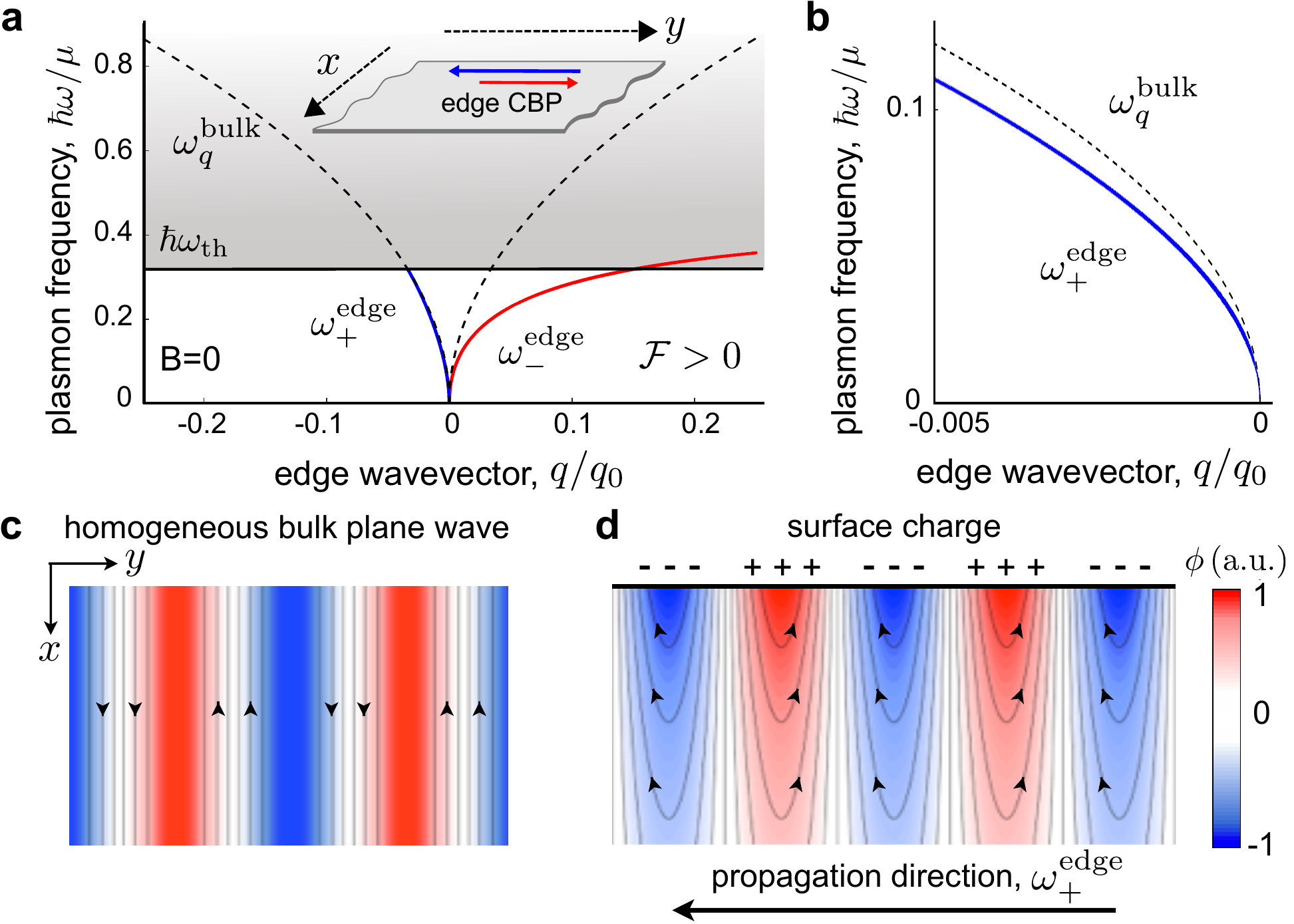}
\vspace{-0.1 in}
\caption{{\bf Chiral Berry plasmons (CBPs) at zero magnetic field. } 
{\bf a)} CBPs are manifested along the edges of a 2D metal with non-vanishing Berry flux, $\mathcal{F}$ (inset). 
Counterpropagating modes exhibit a split dispersion, $\omega_\pm^{\rm edge}(q)$, with the splitting 
{\it increasing} with wave vector $q$ along the edge. 
Above a threshold frequency $\omega_{\rm th}$ (shaded region) the fast mode $\omega_+^{\rm edge}$ merges with bulk plasmon modes, $\omega_{\vec q}^{\rm bulk}$, see Eq.~(\ref{eq:bulk_disp}) and dashed line, leaving a {\it single unidirectional mode} $\omega_-^{\rm edge}$ propagating along the edge. 
{\bf b)} Zoom-in of (a) showing separation of  $\omega_+^{\rm edge}$ and
 $\omega_{\vec q}^{\rm bulk}$ below $\omega_{\rm th}$.  
{\bf c)} and {\bf d)} Electric potential for bulk and edge plasmon modes. 
For bulk modes the anomalous velocity field for electrons (indicated by arrowheads) runs parallel to the wave fronts and does not affect the speed of collective propagation.
Near edges the build-up of surface charges leads to an 
anomalous velocity flow that assists the collective propagation for the $\omega_+^{\rm edge}$ mode, see text below Eq.~(\ref{eq:constitutive}).
For $\omega^{\rm edge}_-$ the anomalous velocity flow opposes the collective motion, see text below Eq.~(\ref{eq:constitutive}).
Parameter values used: $\mathcal{F} = 1.0$ and $a_0=3$ for (a) and (b), see Eq.~(\ref{eq:bulk_disp}), and $|q/q_0|=0.3$, $\mathcal{F} = 0.3$, and $a_0=3$ for (c) and (d).} 
\label{fig:intro}
\vspace{-0.2 in}
\end{figure} 

Here we show that Berry curvature can work in concert with interactions, leading to new types of collective modes in two-dimensional (2D) ``topological'' metals, with
non-vanishing Berry flux (i.e., net Berry curvature), $\mathcal{F}$. In particular, $\mathcal{F}$ gives rise to {\it chiral plasmonic excitations} --- propagating charge density waves with split dispersion for oppositely directed  
modes --- in the absence of a magnetic field (Fig.~\ref{fig:intro}). We refer to these collective modes as ``chiral Berry plasmons'' (CBPs). 
Notably, these chiral modes are localized to the edge of the 2D metal, even in the absence of topological edge states,
and exhibit a rich phenomenology (see below).

We expect CBPs to be manifested in a wide variety of magnetic as well as {\it non-magnetic} materials. The former are materials that exhibit anomalous Hall effects, wherein time reversal symmetry (TRS) breaking is encoded in the Bloch band Berry curvature. The latter include a range of readily available gapped Dirac materials, e.g., transition metal dichalcogenides and graphene/hBN, wherein TRS breaking is achieved by inducing a non-equilibrium
valley polarization~\cite{mak2014}. 
In both cases CBPs are characterized by clear optical signatures such as split peaks in optical absorption.

From a technological perspective, CBPs in non-magnetic materials are particularly appealing since they provide an entirely new platform for achieving a range of magneto-optical effect analogues that are magnetic field-free, and  ``on demand.'' A prime example is optical non-reciprocity~\cite{joannopolus}, which is central for optical device components, e.g., optical isolators and circulators. 
Above a threshold frequency ($\omega_{\rm th}$, see shaded area in 
Fig.~\ref{fig:intro}a), the single unidirectionally propagating mode $\omega^{\rm edge}_-$ allows for chiral transport of light
via coupling to CBPs.
Such CBP mediated waveguides provide a novel paradigm for deep sub wavelength \cite{barnes-plasmons,qplasmonics,polinireview}, linear, and magnetic field free strong non-reciprocity, crucial for miniaturizing optical components. In particular, we predict that CBPs can enable non-reciprocity over a large technologically important bandwidth (THz to mid-IR).

The intrinsic chirality of CBPs starkly contrasts with that achieved via cyclotron motion of charged particles in a magnetic field, with important quantitative as well as qualitative consequences. 
In the latter case, chirality arises via the Lorentz force and gives rise to conventional magnetoplasmons \cite{stormer,fetterexp,evaandrei,magnetoplasmonGAP}. There, the cyclotron frequency, $\hbar \omega_c = \hbar eB/m$ \cite{kohn1961}, determines the constant splitting between magnetoplasmon modes of opposite chirality, which can be of order a few meV for accessible field strengths. In contrast, chirality in CBPs arises from the combined action of plasmonic self-generated electric fields and the anomalous velocity of Bloch electrons, the phase space {\it dual} to the Lorentz force. This combination makes the CBP mode splitting directly sensitive to plasmon wavelength, the Berry flux, and interaction strength, in contrast to magnetoplasmon splittings which only depend on magnetic field\cite{fetterA,fetterexp} \footnote{We note that Kohn's theorem guarantees that splittings are unchanged for parabolic bands; weak renormalizations are expected for linearly dispersing systems such as graphene, where it has been measured to be at most of order 10\%, see e.g., Ref.~\cite{Henriksen}. In contrast, for CBPs the splitting grows directly proportional with interaction strength [see Eq.~(3) of the main text].}. As a result, for short wavelengths and unscreened interactions, large splittings $\hbar \Delta \omega$ of several tens to a hundred meV can be achieved [see Eq.~(\ref{eq:splitting_estimate}) below].

\vspace{2mm}
{\bf Self-Induced Anomalous Velocity -}
The origin of CBPs in two dimensions can be understood 
from the Euler equations for electron density, $n(\vec{r},t)$~\cite{pkbook}: 
\begin{subequations}
\begin{eqnarray}
&&\partial_t n(\vec r, t) + \nabla \cdot \bar{\vec{v}}(\vec{r},t) = 0, \\
&&\partial_t \bar{\vec p}(\vec{r},t) - n(\vec r, t)\, e \nabla \phi(\vec r, t) = 0,
\end{eqnarray}
\label{eq:euler-main}
\end{subequations}
where $\bar{\vec v}(\vec{r},t)$ and $\bar{\vec p}(\vec{r},t)$ are the velocity and momentum density fields, $\phi(\vec r)$ is the scalar electric potential, and $-e<0$ is the electron charge. 
We note that in principle the force equation (\ref{eq:euler-main}b) also includes contributions arising from the stress density of the electronic fluid. However, at long-wavelengths these contributions yield only sub-leading corrections to the plasmon dispersion~\cite{guliani}. As our aim here is to clearly and most simply demonstrate the existence and main features of CBPs, in this work
we neglect small corrections due to the stress density. For an alternative formulation in terms of currents and conductivities, see supplementary online information (SOI)~\cite{appendix}.

In order to fully specify the dynamics, Eq.~(\ref{eq:euler-main}) must be supplemented by a set of {\it constitutive relations} which relate velocity, momentum, density, 
and the electric potential. Plasmons emerge from Eq.~(\ref{eq:euler-main}) as self-sustained collective oscillations of $n(\vec r,t)$, with the potential $\phi(\vec r, t) = \int d^2 \vec r'\, W(\vec r, \vec r') \delta n (\vec r', t)$ generated by the plasmon's density fluctuations
$\delta n = n(\vec r, t) - n_0$. Here $n_0$ is the average carrier density, and $W(\vec r, \vec r')$ is the Coulomb interaction. 

As we argue, in the presence of Berry curvature, the constitutive relations take on an anomalous character.
This can be seen by starting with the quasiparticle semiclassical equations of motion~\cite{dixiao}, $\vec v_{\vec p} = \frac{\partial \epsilon_{\vec p}}{\partial \vec p} + \frac{1}{\hbar}\dot{\vec p} \times \boldsymbol{\Omega} (\vec p)$,  $\dot{\vec{p}}  = e \nabla \phi(\vec r)$, where $\vec v_{\vec p} = \dot{\vec{r}}$ is the quasiparticle velocity and $\epsilon_{\vec p}$ and $\boldsymbol{\Omega}(\vec{p}) = \Omega(\vec p) \hat{\vec{z}}$ are the Bloch band dispersion and Berry curvature, respectively~\footnote{The Berry curvature $\Omega(\vec p) = \nabla_{\vec k} \times \vec A_\vec{k}$ depends on the crystalline Bloch wavefunctions $|u_{\vec k}\ra$, where $\vec A_\vec{k} = \la u_\vec{k} | i \nabla_{\vec k} | u_\vec{k}\ra$ is the Berry connection.}. 
The corresponding velocity density 
fields are found from these relations and the phase space density $f_i(\vec{r},\vec{p},t)$ by summing over all momentum $\vec{p}$ and bands $\{i\}$, $\bar{\mathcal{O}}(\vec{r},t) =\sum_{i}\int d^2 \vec p\, \mathcal{O}f_i(\vec{r},\vec{p},t)/ (2\pi \hbar)^{2} $. 
 This gives:
\begin{equation}
\bar{\vec v}(\vec{r},t) = \frac{\bar{\vec p}(\vec{r},t)}{m} + \bar{\vec v}_a(\vec{r},t), \ \ \bar{\vec v}_a = \frac{e\mathcal{F}}{\hbar} \Big[ (\nabla \phi) \times \hat{\vec z} \Big],
\label{eq:constitutive}
\end{equation}
where $\mathcal{F} = \sum_{i} \int d^2 \vec p\, \Omega_i (\vec p) f_{i}^0(\vec p)/ (2\pi \hbar)^2$ 
is the (dimensionless) Berry flux, with $ f_{i}^0(\vec p)$ the equilibrium band occupancy; here we have only kept terms linear in $\delta n$ and $\nabla \phi$.
In addition to the conventional first term, which governs the behavior of ordinary plasmons, $\bar{\vec v}(\vec{r},t)$ in Eq.~(\ref{eq:constitutive}) exhibits a self-induced anomalous velocity component $\bar{\vec{v}}_a$ that yields chirality as shown in Fig.~\ref{fig:intro}.
Note that the mass $m$ appearing in Eq.~(\ref{eq:constitutive}) is the {\it plasmon mass}, which characterizes the collective motion of the Fermi sea~\cite{donheeham}.

CBP chirality can be understood intuitively by examining the anomalous velocity pattern set up by the plasmon's electrostatic potential $\phi$ (a more complete treatment is given below).
Due to the cross product in Eq.~(\ref{eq:constitutive}), the anomalous velocity flow is directed along the equipotential contour lines of $\phi$ (see arrowheads in Figs.~\ref{fig:intro}c and \ref{fig:intro}d). Near an edge, surface charges associated with the plasmon wave produce a potential as shown in Fig.~\ref{fig:intro}d. 
The corresponding anomalous velocity field directs electrons into the nodal regions to the {\it left}
of each region of negative charge build up (i.e., excess electron density), for the orientation shown and $\mathcal{F} > 0$. 
Thus for a leftward direction of plasmon propagation, the anomalous velocity flow {\it assists} the collective motion  of the electronic density wave, leading to faster propagation $\omega_+^{\rm edge}$. For the right-propagating mode the anomalous flow works against the collective motion, yielding slower propagation $\omega^{\rm edge}_-$.

Crucially, 
 $\bar{\vec{v}}_a$ depends directly on the self-generated electric field  
$-\nabla \phi(\vec r)$.
Consequently, the magnitude of the splitting 
is governed by the wave vector $q$ and the strength of Coulomb interactions. 
We emphasize, however, that CBPs are a {\it linear} phenomenon, with the mode splitting $\Delta \omega$ independent of the magnitude of $\phi$.  As we will show, the $q$-dependent CBP splitting can be large:
\be
\hbar \Delta \omega = \hbar(\omega^{\rm edge}_+ - \omega^{\rm edge}_-) 
\approx  \mathcal{A} \frac{e^2 }{\kappa} \mathcal{F} |q|,
\label{eq:splitting_estimate}
\ee
where $\mathcal{A}$ is a numerical prefactor of order unity that depends on geometry, and we have used the 2D Coulomb potential $W(\vec q) = 2\pi e/ ( \kappa|\vec q|)$ with background dielectric constant $\kappa$. 
For edge CBPs, we find $\mathcal{A} = 8 \sqrt{2} \pi/9 $ (see below), yielding large splittings 
$\hbar\Delta \omega \approx 6 -60 \, {\rm meV}$ for $q= 1 -10 \, \mu {\rm m}^{-1}$ (here we have used $\mathcal{F}= 1$, $\kappa =1$).
The appearance of $e^2/\kappa$ on the right hand side of Eq.~(\ref{eq:splitting_estimate}) signals the crucial role interactions play in $\Delta \omega$.

\vspace{2mm}
{\bf Chiral Edge Berry Plasmons -}  
We now analyze collective motion 
described by Eq.~(\ref{eq:euler-main}), treating the electric potential $\phi$ self-consistently. For an infinite bulk, applying $\partial_t$ to the continuity equation (\ref{eq:euler-main}a), using Eq.~(\ref{eq:constitutive}), and substituting in the force equation (\ref{eq:euler-main}b) yields 
\be
- \partial_t^2 \delta n = \frac{n_0e}{m}\nabla^2 \phi , \ \ \phi(\vec r) = \int_{-\infty}^{\infty}\!\!\! d\vec r'\, W (\vec r- \vec r')\, \delta n(\vec r').
\label{eq:reduced}
\ee
In arriving at Eq.~(\ref{eq:reduced}), we have used $\nabla \cdot \bar{\vec v}_a  \propto \partial_x\partial_y \phi(\vec r)  -  \partial_y \partial_x \phi(\vec r) =0$. Importantly, Berry flux $\mathcal{F}$ is {\it absent} in 
Eq.~(\ref{eq:reduced}) and has no effect on bulk plasmon dispersion. 
Indeed, decomposing into Fourier modes $\delta n \sim e^{i\omega t - i\vec q \cdot \vec r}$ 
and using the Coulomb interaction $W(\vec r, \vec r')=-e/(\kappa|\vec{r}-\vec{r}'|)$ yields the usual 2D plasmon dispersion 
\be
\label{eq:bulk_disp} (\hbar \omega_{\vec q}^{\rm bulk} /\mu )^2= a_0 (q/q_0),\ \ a_0 = \frac{2\pi \hbar^2}{m\mu} n_0,\ \  q_0 = \frac{\kappa \mu}{e^2}, 
\ee
which remains gapless at $q = 0$.  In contrast, bulk magnetoplasmons are {\it gapped} due to cyclotron motion~\cite{magnetoplasmonGAP,fetterexp,evaandrei}.

\begin{figure}[t]
\includegraphics[width=\columnwidth]{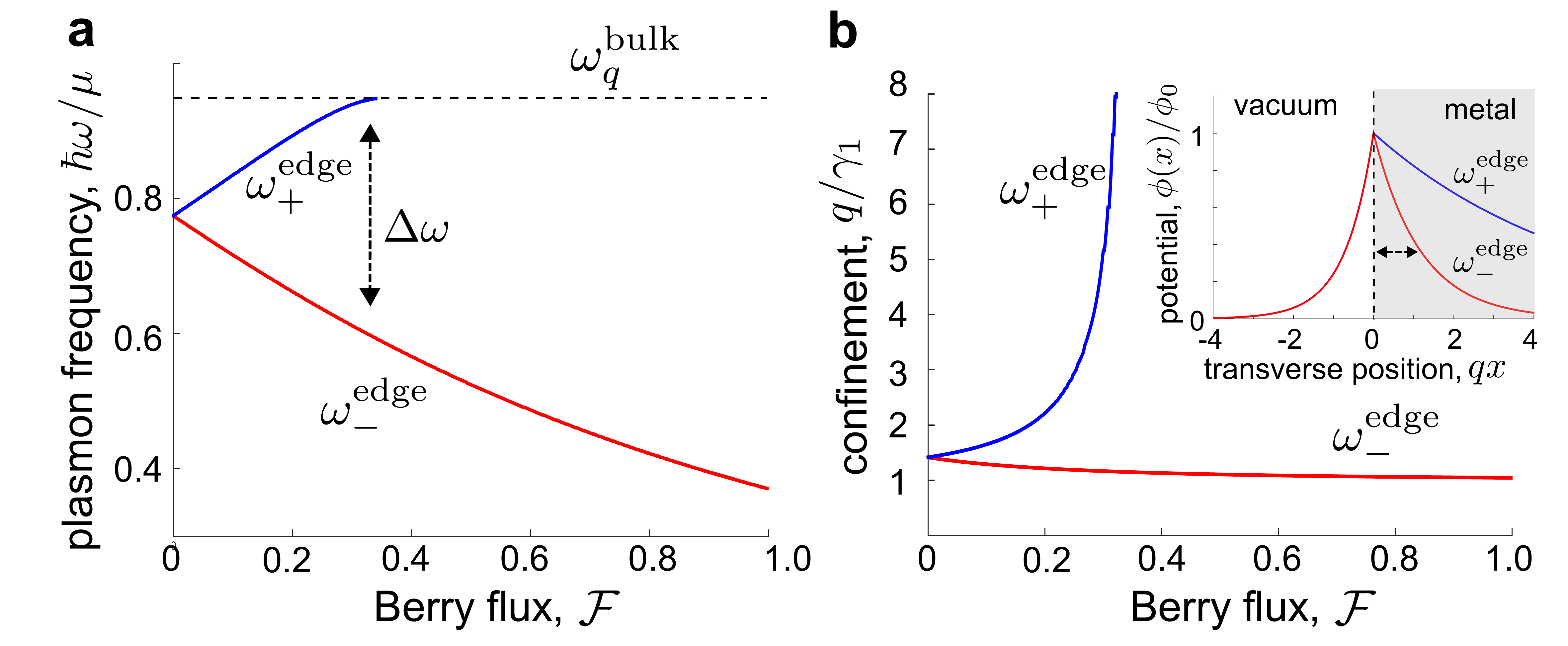}
\caption{{\bf Chiral Berry plasmons at the boundary of a half plane.} {\bf a)} The CBP 
frequency splitting increases with increasing Berry flux, $\mathcal{F}$. Parameter values are as in Fig.~\ref{fig:intro}, but with $|q|/q_0 = 0.3$. {\bf b)} CBP edge-confinement length, $\gamma_1^{-1}$, of $\omega^{\rm edge}_+$ and $\omega^{\rm edge}_-$, see Eq.~(\ref{eq:edge}), for increasing $\mathcal{F}$. For large $\mathcal{F}$, the confinement length of the $\omega^{\rm edge}_+$ mode diverges, indicating that it joins the bulk, while the $\omega^{\rm edge}_-$ mode becomes more confined to the edge. (inset) EM profile shown for $|\mathcal{F}| = 0.3$, exhibiting confinement of $\omega^{\rm edge}_\pm$ to the edge.}
\label{fig2}
\end{figure} 

Close to a boundary, the situation is dramatically altered: Berry flux $\mathcal{F}$ leads to the emergence of 
{\it chiral edge plasmons}, one-dimensional chiral analogues of surface plasmons \cite{fetterA,fetterB}. 
This is most easily illustrated for
an infinite metallic half-plane, 
where $n(\vec r, t)$ and $\bar{\vec v} (\vec r, t)$ are generically finite for $x\geq 0$, but are zero for $x<0$. 
Here the plasmon propagates as a plane wave along $y$, 
with the fields in Eq.~(\ref{eq:reduced}) taking the form 
\begin{equation}
\label{eq:plane_waves}
\phi(\vec r, t) = \phi_q(x) e^{ i\omega t-iqy}, \ \delta n (\vec r, t) = \delta n_q (x) e^{ i\omega t-iqy}.
\end{equation} 

The presence of the edge allows charges to accumulate, see Fig.~\ref{fig:intro}(d). 
Indeed, inserting the fields defined above into Eq.~(\ref{eq:euler-main}a) yields $\delta n(\vec r) e^{i\omega t}= -\nabla \cdot \big[\bar{\vec v}(\vec r,t) \Theta (x) \big]/i\omega$. 
Here we have explicitly inserted $\Theta(x)$ to emphasize the 
vanishing of the velocity density outside the metallic half-plane [$\Theta(x) =1$ for $x\geq 0$ and $\Theta(x) = 0$ for $x<0$]. 
Due to the $\Theta(x)$ inside the divergence above, a non-zero $\bar {\vec v}_x$  flowing into the boundary induces an oscillating surface charge component of $\delta n(\vec r) e^{i\omega t}$ concentrated at the edge $x=0$~\cite{fetterB,Mikhailov}. By replacing $\delta n (\vec{r},t)$ by $-\nabla \cdot \big[\bar{\vec v}(\vec r,t) \Theta (x) \big]/i\omega$ in the integral for $\phi (\vec r,t)$ in Eq.~(\ref{eq:reduced}), we thus find a jump 
condition for $\partial_x \phi_q(x)$ at $x=0$:
\be
\partial_x \phi_q\big|_{0^+}  -  \partial_x \phi_q \big|_{0^-} = \frac{1}{i\omega} \big (\partial_x W_q\big|_{0^-} -\partial_x W_q \big|_{0^+}   \big)\bar{\vec v}_x |_{0^+},
\label{eq:velbc}
\ee
where $W_q(x) = -(e/\kappa ) \int_{- \infty}^{\infty} dk\, e^{ikx}/|q^2 + k^2|^{1/2}$ is the 1D effective Coulomb interaction.
Here we decomposed Eq.~(\ref{eq:reduced}) into plane-waves using Eq.~(\ref{eq:plane_waves}), and integrated across the delta function, $\partial_x\Theta(x) = \delta(x)$, which accounts for the surface charge layer. 
The parameter $q$ corresponds to the wavevector in $y$ (along the edge),  
and $k$ describes variations in $x$ (perpendicular to the edge).
Crucially, finite $\mathcal{F}$ in Eq.~(\ref{eq:constitutive}) makes $\bar{\vec v}_x|_{0^+}$
depend on both the {\it magnitude} and {\it sign} of $q$ along the edge~\footnote{The jump discontinuity boundary condition in Eq.~(\ref{eq:velbc}) depends on the perpendicular velocity density at the boundary, $\bar{\vec{v}}_x(x = 0)$.  We use the subscript $0_+$ in the text to emphasize that the velocity density should be evaluated on the metallic side, $x \ge 0$.}. 

We now find edge CBP solutions of Eq.~(\ref{eq:reduced}) using the boundary condition (\ref{eq:velbc}) and continuity of $\phi(\vec r)$.  
This problem is a 
non-local integro-differential equation owing to the kernel $W_q(x)$. 
We adopt a simplified approach, 
approximating $W_q(x)$ by 
a similar kernel $\tilde{W}_q(x)$ with the same area and second moment~\cite{fetterA,fetterB}: $\tilde{W}_q(x) = -\frac{e}{\kappa}\int_{-\infty}^{\infty} dk\, 2qe^{ikx} /(k^2 + 2q^2)$. 
We emphasize that $\tilde{W}_q(x)$ is extended in $x$, and captures the long-range Coulomb behavior of $W_q(x)$. Indeed, the Fourier transforms of $W_q(x)$ and $\tilde{W}_q(x)$ match for small $k/q$. This method has been used successfully to mimic Coulomb interactions in isolated systems
~\cite{fetterexp,wang11}. 

Crucially, the simple form of $\tilde{W}_q(x)$ above allows the integro-differential equation in Eq.~(\ref{eq:reduced}) to be expressed as a purely differential one with $\phi_q(x)$ obeying
\be
(\partial_x^2 - 2q^2) \phi_{q}^{>} (x) =  \frac{4 \pi e |q|}{\kappa} \delta n_q (x), \ \ (\partial_x^2 - 2q^2) \phi_{q}^{<} (x) =  0,
\label{eq:poisson}
\ee
where $\phi_q^>$ and $\phi_q^<$ are defined inside ($x \ge 0$) and outside ($x < 0$) the sample, respectively. 
Eq.~(\ref{eq:poisson}) yields simple $\phi$ profiles 
\be
\phi_q^< (x)= \phi_0\, e^{\gamma_0x}, \quad \phi_q^> (x)= \phi_1\, e^{-\gamma_1 x}, 
\label{eq:edge}
\ee 
where $\gamma_{0}= \sqrt{2} |q|$, and $\gamma_{1} = \sqrt{2} |q|\, \{[ (\omega^{\rm bulk}_\vec q)^2 - \omega^2]/[2(\omega^{\rm bulk}_\vec q)^2 - \omega^2] \}^{1/2}$. 
The latter was obtained 
from Eqs.~(\ref{eq:reduced}) and (\ref{eq:poisson}) by eliminating $\delta n_q$.

Using Eq.~(\ref{eq:edge}), the boundary conditions of continuous $\phi$ and the jump condition (\ref{eq:velbc}) can be expressed compactly via the relation $\mathcal{S} \Phi = 0$, with $\Phi = (\phi_0, \phi_1)^T$:
\be
\mathcal{S} = \left( \begin{array}{cc}  1 & -1 \\ \sqrt{2}|q| & \gamma_1 - D 
\end{array} \right), \quad D = 
\frac{2\gamma_1(\omega^{\rm bulk}_q)^2}{\omega^2}- \frac{ q^2\tilde{ \mathcal{F}}{\rm sgn}(q)}{\omega},
\label{eq:scattering}
\ee
where $D$ was obtained using Eqs.~(\ref{eq:constitutive}) and (\ref{eq:velbc}), and $\tilde{F} = 4\pi e^2 \mathcal{F} /\kappa \hbar$. 
Left- and right- moving plane wave modes along the edge, $\omega_\pm^{\rm edge}$, can be identified through the zero modes of $\mathcal{S}$. We first note that for $\tilde{\mathcal{F}}=0$, solving ${\rm det}(\mathcal{S}) =0$ yields degenerate non-chiral edge modes with $\omega_\pm^{\rm edge} = (2/3)^{1/2} \omega_{\vec q}^{\rm bulk}$, see Refs.~\cite{fetterA,fetterB,fetterexp}.

For non-zero $\tilde{\mathcal{F}}$, the zero mode solutions $\omega_\pm^{\rm edge}$ of Eq.~(\ref{eq:scattering}) become split, yielding chiral edge plasmons (CBPs) as shown in Figs.~\ref{fig:intro} and \ref{fig2}. The modes $\omega_+^{\rm edge}$ and $\omega_-^{\rm edge}$ propagate as waves in opposite directions along the edge, with faster and slower speeds, respectively. Importantly, the frequencies $\omega_\pm^{\rm edge}$ depend both on $q$ along the edge (Fig.~\ref{fig:intro}a), as well as $\mathcal{F}$ (Fig.~\ref{fig2}a). The splitting between modes grows with $q$ and $\tilde{\mathcal{F}}$, since $\bar{\vec v}_a = e\nabla \phi \times \mathcal{F}/\hbar$ [see Eq.~(\ref{eq:constitutive})]. Indeed, for small $q\tilde{\mathcal{F}}$ (so that $q \tilde{\mathcal{F}} \ll \omega_q^{\rm bulk}$), we obtain an approximate dispersion for edge CBPs as
$
 \omega_\pm^{\rm edge} \approx  (2/3)^{1/2} \omega^{\rm bulk}_\vec q \pm \sqrt{2}|q| \tilde{\mathcal{F}}/9 + \mathcal{O}(q^2 \tilde{\mathcal{F}}^2)
 $. 
As a result, we obtain the $q$-dependent $\Delta \omega$ in Eq.~(\ref{eq:splitting_estimate}). 
This behavior sharply contrasts with
that of magnetoplasmons, which have a $q$-independent splitting 
given by the cyclotron frequency~\cite{kohn1961,fetterA,fetterB,fetterexp}.  
As a result, far larger splittings, arising from interactions, can be achieved for CBPs. 

Interestingly, for large enough $q$ and/or $\mathcal{F}$, 
the $\omega_+^{\rm edge}$ mode (blue line in Figs.~\ref{fig:intro}a,b and \ref{fig2}a) merges with the bulk plasmon mode $ \omega_\vec{q}^{\rm bulk}$ (dashed line). As this mode merges with the bulk, its potential profile ceases to be localized along the edge.
This is shown by a diverging confinement length of the electric potential, $\gamma_{1}^{-1}$, see Fig.~\ref{fig2}b. 
In contrast, $\omega_-^{\rm edge}$ stays far from the bulk dispersion, yielding a potential (and electric field)  
tightly confined to the edge. The threshold $\omega_{\rm th}$ above which $\omega_+^{\rm edge}$ merges with the bulk can be obtained from Eq.~(\ref{eq:scattering}). Setting $\omega = \omega_\vec q^{\rm bulk}$ in Eq.~(\ref{eq:scattering}) yields the threshold frequency 
\be
\hbar \omega_{\rm th} = \frac{\hbar^2 n_0}{\sqrt{2}m|\mathcal{F}|} = 18.3 \,  \frac{n_0 [{\rm cm}^{-2}]/10^{12}}{(m [m_e]/0.03) \times |\mathcal{F}|}\, {\rm meV},
\ee
with $m_e$ the free electron mass. 
For scale we consider a plasmon mass $m\!\sim 0.03\, m_e$, as measured in graphene~\cite{donheeham}. 

Conservation of $\omega$ and $q$ along the edge 
protect the $\omega_-^{\rm edge}$ mode from coupling to bulk 2D plasmons.
Scattering processes that relax $q$ contribute to propagation losses. However, the tight edge confinement of the $\omega_-^{\rm edge}$ mode  
suppresses its electric field in the bulk regions (see Fig.~\ref{fig2}b inset), 
suppressing its coupling to bulk plasmons.
Therefore, above the threshold $\omega_{\rm th}$ (gray region in Fig.~\ref{fig:intro}a), the single, well-defined, 
$\omega_-^{\rm edge}$ mode propagates unidirectionally along the edge.
When hybridized with light, it will allow for strong non-reciprocal propagation of CBP-polaritons without magnetic field (see below).

\vspace{2mm} 
{\bf Experimental Signatures of CBPs -} 
Strong plasmon mediated light-matter interactions~\cite{qplasmonics,polinireview,barnes-plasmons} make optics an ideal means of probing/controlling CBPs. Photon coupling to plasmons with gapless dispersion (e.g., 2D plasmons, and CBPs here) can be achieved through strategies such as gratings, and prism geometries~\cite{barnes-plasmons}. Observing unidirectional (non-reciprocal) propagation in such setups can reveal the existence of CBPs. For demonstration, we detail an alternative experimental probe: CBP-photon coupling in finite geometries, such as metallic disks, where dipolar plasmonic modes can dominate optical absorption~\cite{barnes-plasmons,stormer}. 

\begin{figure}[t]
\includegraphics[width=\columnwidth]{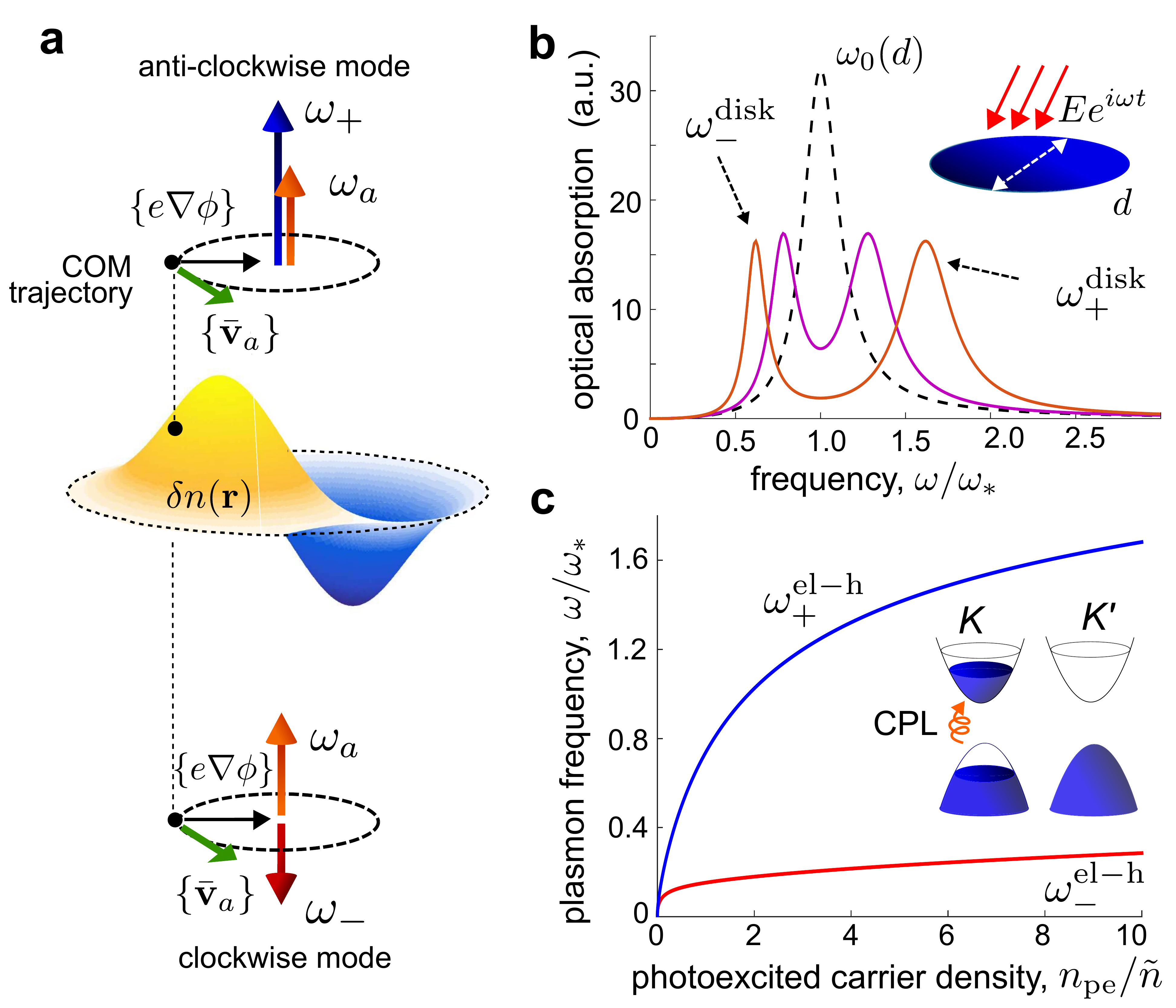}
\caption{{\bf Chiral Berry plasmons in a disk and in valley polarized gapped Dirac systems.} {\bf a)} Illustration of CBPs in a disk, showing anti-clockwise/clockwise rotating dipole modes. The mode splitting arises from the intrinsic angular frequency, $\omega_{a}$,  
induced by the combination of $\mathcal{F}$ and the self-induced electric field. {\bf b)} Optical absorption split peaks for dipolar CBPs in a disk (panel a) obtained by inverting Eq.~(\ref{eq:eomdisk}) for $\mathcal{F} = 0,0.5,1.0$ (dashed black, purple, orange). Smaller disk sizes and/or larger $\mathcal{F}$ produce larger splittings,  
see Eq.~(\ref{eq:dispersion}). Parameter values: $\omega_0/\omega_* = 1$ and $\Gamma/\omega_* = 0.25$.  {\bf c)} CBP dipole modes for valley polarized gapped Dirac systems, with $n_{K} = n_{\rm pe}$, and $n_{K'} =0$. Characteristic density is $\tilde{n} = \Delta^2/{4\pi v_F^2 \hbar^2}$. 
(inset) Valley polarization may
be induced by above-gap circularly polarized light.} 
\label{fig3}
\end{figure}

In metallic disks with finite $\mathcal{F}$,  
CBPs manifest as clockwise/anti-clockwise moving plasmonic dipole modes (Fig.~\ref{fig3}a).
These modes can be described via a simple oscillator model for the motion of the dipolar CBP center of mass (COM) coordinate~\cite{appendix}, 
$\{ \vec x\}$, where $\{ \cdot \}$ denotes the COM average. Here $\{ {\vec v}_a \} \approx \mathcal{F}\,\{e\nabla \phi \}  \times \hat{\vec z}$ (green arrow) gives rise to an intrinsic angular frequency $\omega_{a}$ of plasmons in a disk (orange arrow), which adds to (subtracts from) the plasmon frequency $\omega_0$ to produce non-degenerate anti-clockwise (clockwise) rotating modes (Fig.~\ref{fig3}a); see detailed derivation in SOI. Here we have used $\mathcal{F}$ pointing to positive $\hat{\vec{z}}$. 
A bosonic analogue for ultra cold atomic gases is discussed in Ref.~\cite{cooper}.

With an a.c.~probing electric field $\sim \vec{E}\, e^{i\omega t}$,  
the COM equations of motion are:  
\be
\begin{array}{rcl}
  \partial^2_t\{{x}\} + \omega_0^2\, \{x\} + \omega_a \partial_t\{{y}\} &=\! -e E_x\, e^{i\omega t}\\[0.3em]
  \partial^2_t\{{y}\} + \omega_0^2\, \{y\} - \omega_a \partial_t\{{x}\} &=\! -eE_y\, e^{i\omega t},
\end{array}\, \omega_{a}= \frac{\mathcal{F} \omega_0^2 m}{n_0 \hbar}.
\label{eq:eomdisk}
\ee
Here $\omega_0 (d)$ is the bare plasmon frequency in a disk of diameter $d$, in the absence of Berry curvature. 

Writing the current density as $\vec j = e n_0 \, \partial_ t \{ \vec x \}$, we invert the COM equations of motion 
to obtain the optical absorption (real part of the longitudinal conductivity~\cite{appendix}).
As shown in Fig.~\ref{fig3}b, we find a split peak structure with the dipolar CBP peaks given by the 
poles of Eq.~(\ref{eq:eomdisk}):
\be
\omega_\pm^{\rm disk}= \sqrt{\omega_0^2 + \frac{\omega_{a}^2}{4}} \pm \frac{\omega_{a}}{2}, \ 
\hbar\Delta \omega \equiv \mathcal{F} \hbar \omega_* \approx \frac{9 \mathcal{F}}{d [\mu{\rm m}]} \, {\rm meV},
\label{eq:dispersion}
\ee
where $\Delta \omega = \omega_+^{\rm disk} - \omega_-^{\rm disk} = \omega_{a}$, and $\hbar \omega_* =  \omega_0^2 m/n_0$. 
On the right side, we have estimated $\omega_0^2 \approx 2\pi e^2 n_0 |\vec q| / m$, with $|\vec q| \approx 1/ d$ (approximating the lowest lying plasmonic excitation in a disk)~\footnote{Comparing this estimate to plasmon frequencies obtained in graphene disks (e.g.,~in Ref.~\cite{avouris}) we obtain zero field plasmon frequencies in the disk geometry to within a factor of unity. Using parameters reported in Ref.~\cite{avouris}, our estimate yields $\hbar\omega_0 \approx \sqrt{2\pi e^2 n_0 / (m d)} \approx 24 \, {\rm meV}$ for a disk of $d=3 \, \mu {\rm m}$ and reported doping $\mu = -0.54 \, {\rm eV}$. Ref.~\cite{avouris} observed a zero field plasmon resonance at $\omega_0 = 130 \, {\rm cm}^{-1} = 16 \, {\rm meV}$.}.  Here we have used $\kappa =1$. Importantly, $\Delta \omega$ depends on the disk size, $d$, a unique property of CBPs. 

The tunable optical absorption split peak structure (via $\mathcal{F}$ and $d$) in the absence of an applied magnetic field gives a clear experimental signature of CBPs. 
In plotting Fig.~\ref{fig3}b, we have included the damping rate phenomenologically via $\partial_t^2 \to \partial_t^2 + \Gamma \partial_t$, yielding a Lorentzian lineshape with its half-width determined by $\Gamma$. 
Split peaks are clearly visible when $\Delta \omega \gtrsim \Gamma$, yielding peaks at $\omega_\pm^{\rm disk}$. To give a sense of scale, 
we note a typical value $\hbar \Gamma \sim {\rm few}\,\, {\rm meV}$, see e.g.~Ref.~\cite{avouris} where $\hbar \Gamma \approx 4 \,{\rm meV}$ was measured in graphene disks. 
Using $\Gamma/ \omega_* = 0.25$ and taking $\mathcal{F} = 1$, clearly resolved $\omega_\pm^{\rm disk}$ peaks can be resolved for disk sizes $d \lesssim 1 \, \mu {\rm m}$ (Fig.~\ref{fig3}b).

\vspace{2mm}
{\bf CBPs in Anomalous Hall Materials - }
We now discuss 
materials where CBPs can be realized.
We predict that metallic systems with non-vanishing 
$\mathcal{F}$ will generically host CBPs. Finite $\mathcal{F}$ 
requires broken time reversal symmetry, and may arise in magnetically ordered systems 
or out-of-equilibrium non-magnetic systems (see below). The former includes magnetically doped semiconducting quantum wells (see e.g., Ref.~\cite{culcer}, where $\mathcal{F}\approx 1/2$ was predicted) and topological insulators~\cite{yu2010,nagaosa2011,qahe,wang14}.

As a concrete example, we examine the 
magnetically doped topological insulator chromium doped thin-film (Bi,Sb)$_2$Te$_3$, which was recently experimentally realized~\cite{qahe}.
When moderately doped with electrons or holes, 
we estimate that $\mathcal{F} \sim 1$  can be achieved based on the measured anomalous Hall conductivity $\sigma_{xy}^{B=0} \lesssim e^2/h$ \cite{qahe}.
This yields a large splitting $\hbar \Delta \omega\approx 10-100 \, {\rm meV}$ for short wavelength plasmons, see Eqs.~(\ref{eq:splitting_estimate}) and (\ref{eq:dispersion}). When probed in the disk geometry, we predict this system will yield two split absorption peaks in the absence of magnetic field.

\vspace{2mm}
{\bf ``On-demand'' CBPs in Non-Magnetic Materials - } 
Intriguingly, finite $\mathcal{F}$ can also be achieved in {\it non-magnetic} materials, without an applied magnetic field. This includes, for example,
gapped Dirac materials where inversion symmetry is broken. 
In these, a valley polarization (Fig.~\ref{fig3}c inset) can be induced by circularly polarized light~\cite{yao08}, yielding an anomalous Hall effect that  
has recently been observed~\cite{mak2014}; see materials discussion below.

Can CBPs exist in
photoexcited systems? 
To analyze this, we focus on nominally time reversal invariant gapped Dirac materials.
We model the valley dependent $\mathcal{F}$ as~\cite{lensky}:  
$\mathcal{F}_{K,K'}=  \tau_z N_s \frac{\tilde{n}^{1/2}}{ 2({n_{K,K'} + \tilde{n}})^{1/2}}\, {\rm sgn\, \Delta}$, where $n_{K,K'}$ are the valley carrier densities, $\tau_z = \mp$ for the $K,K'$ valleys, $N_s = 2$ is the spin degeneracy, and $\tilde{n} = \Delta^2/{4\pi v_F^2 \hbar^2}$ gives a characteristic density scale. The bandgap is $2|\Delta|$. 
When $n_K= n_{K'}$, as in equilibrium, the total flux $\mathcal{F} = \mathcal{F}_K + \mathcal{F}_K'$  
vanishes. 

Interestingly, when the system is pushed out of equilibrium, e.g., by circularly polarized light, the populations in the valleys may become imbalanced, $n_K\neq n_{K'}$ \cite{yao08,appendix} (Fig.~\ref{fig3}c inset).  As a result, the net Berry flux for the entire electronic system (summed over both valleys) is {\it non-zero}, yielding $\mathcal{F}_{\rm pe} \neq 0$. 
We analyze the collective modes of the photoexcited electron-hole system 
 in the disk geometry following Eq.~(\ref{eq:eomdisk}) above, accounting for the mutually attracting electron and hole populations and using $\mathcal{F} = \mathcal{F}_{\rm pe}$. This yields two chiral CBP modes, $\omega_\pm^{\rm el-h} = \sqrt{\omega_{a}^2 + 2\omega_0^2} \pm \omega_{a}$, see SOI. Estimating $\omega_0$ as below Eq.~(\ref{eq:dispersion}), we obtain the $\omega_\pm^{\rm el-h}$ curves in Fig.~\ref{fig3}c with $\omega_*$ given by Eq.~(\ref{eq:dispersion}). 
The frequencies of both modes
vanish at zero photoexcited carrier density, $n_{\rm pe}$.
The mode splitting increases with 
$n_{\rm pe}$, and can reach sizable values for $n_{\rm pe} \geq \tilde{n}$, reflecting $\mathcal{F}_{\rm pe}$ pumping.

Large splittings require
$n_{\rm pe} \gtrsim \tilde{n}$. The large gaps $\Delta \gtrsim 1\, {\rm eV}$ of many transition metal dichalcogenides~\cite{mak2014, jones, heinz_mote2} yield large characteristic densities requirements, $ \tilde{n} \gtrsim 10^{13}\, {\rm cm}^{-2}$. In contrast, other gapped Dirac materials such as G/hBN~\cite{hunt13,woods14,gorbachev14} and dual-gated bilayer graphene~\cite{bilayergraphene} possess $\Delta \approx 10-200 \, {\rm meV}$, yielding significantly smaller and more favorable $\tilde{n}$. These materials possess valley-selective optical selection rules~\cite{yao08,appendix}, and present an ideal venue to achieve maximal $\mathcal{F} = 1$ (and large $\Delta \omega$), even with relatively weak pump fluence~\footnote{Note that the (neutral) excitons that may form in some large-gap materials feel zero net Berry curvature~\cite{yao14}, and to first approximation do not contribute to CBPs.}.

A further promising strategy to achieve large CBP splittings is to stack $m$ 
layers of gapped Dirac materials 
on top of each other,  
with no tunnel coupling between 
the layers. Stacking achieves (i) larger photo excited carrier densities due to the increased absorption, and (ii) a larger maximal Berry flux (when $n_{\rm pe} \gg \tilde{n}$) and, hence, larger CBP splittings; the maximal Berry flux is $\mathcal{F}_{\rm max} = m \mathcal{F}_{\rm single}$. In such a structure, the long-range  
Coulomb interaction allows the photoexcited densities in different layers to oscillate collectively.

{\bf Conclusion - } CBPs are robust collective excitations of metallic systems, arising from two simple ingredients: Berry flux and interactions.  
Our analysis indicates that CBPs survive for both weak and strong interactions.
As a result, we conclude that CBPs are generic in 
metallic anomalous Hall phases, including out-of-equilibrium states where finite $\mathcal{F}$ 
emerges from driving (e.g., in optically pumped valley polarized gapped Dirac materials). 
Indeed, this feature allows CBPs to be used for all-optical diagnostics of anomalous Hall and topological phases, as well as pumped or periodically driven systems, e.g., as in Ref.~\cite{wangyao07, kitagawa}.
Optical probes of the latter are particularly appealing since transport measurements require contacts that often complicate and destroy the novel electron distributions and coherences of driven systems. 

Perhaps the most appealing prospect is coupling CBPs with light for sub-wavelength, and strong non-reciprocal propagation. Exhibiting a single chiral mode at large $q$ (large splitting, and large frequency), precisely where plasmons give large compression of optical mode volume, hybrid CBP-polaritons are strongly non-reciprocal~\cite{joannopolus}. As we propose, 
CBP mediated unidirectional waveguides can be realized in readily available {\it non-magnetic} materials (e.g., the van der Waals material class). CBP based non-reciprocity, if realized experimentally,
stands to play a vital role in the miniaturization of optical components that are magnetic field-free. 

{\it Note added:}
During the review of our manuscript, a related work on CBPs in photoexcited gapped Dirac systems appeared, see Ref.~\cite{kumar}.

{\bf Acknowledgements - }
We thank J. Alicea, J. Eisenstein, M. Kats, L. Levitov, M. Schecter, and B. Skinner for useful conversations.  
JS acknowledges support from the Walter Burke Institute for Theoretical Physics as part of a Burke fellowship at Caltech. MR acknowledges support from the Villum Foundation and the People Programme (Marie Curie Actions) of the European Union's Seventh Framework Programme (FP7/2007-2013) under REA grant agreement PIIF-GA-2013-627838.

\newpage

\section{Supplementary Online Information for ``Chiral plasmons without magnetic field''}

In this supplement we include discussions of an alternative formulation of chiral Berry plasmons based on current densities and the conductivity tensor, chiral Berry plasmon dipole modes in a disk, optical selection rules for gapped Dirac materials with inversion symmetry breaking, and photo-induced chiral Berry plasmons.

\setcounter{equation}{0}
\renewcommand{\theequation}{S-\arabic{equation}}
\makeatletter
\renewcommand\@biblabel[1]{S#1.}

\subsection{Current density and the conductivity tensor}

For a complementary view to the approach employed in the main text, here we present an alternative formulation of the plasmon equations of motion.
Rather than focusing on the semiclassical equations of motion for the particle and velocity density fields, here we work with the charge density $\rho(\vec{r},t) = -en(\vec{r},t)$, the current density $\vec{j}(\vec r, t)$, and the conductivity tensor $\boldsymbol{\sigma}$. 
As we will be interested in oscillatory collective (plasmon) modes, it will be useful to work in terms of Fourier modes, $\rho(\vec{r},t) = \rho_\omega(\vec{r})e^{i\omega t}$, $\vec{j}(\vec{r},t) = \vec{j}_\omega(\vec{r}) e^{i\omega t}$, and $\phi(\vec{r},t) = \phi_\omega(\vec{r})e^{i\omega t}$.
The charge and current densities obey a continuity equation, 
which in terms of the Fourier modes reads:
 \be
 i\omega \rho_\omega (\vec r) + \nabla \cdot \vec j_\omega (\vec{r}) = 0,\ \  \vec j_\omega(\vec r) = \bm{\sigma}(\omega) \nabla \phi_\omega(\vec r). 
 \label{eq:currentcon}
\ee
Importantly, even at $B = 0$, the conductivity $\boldsymbol{\sigma}(\omega)$ possesses off-diagonal contributions, captured by the anomalous Hall conductivity $\sigma_{xy}^{\rm AH}$.

In terms of frequency dependence, the longitudinal conductivity takes a standard Drude form, while the Hall conductivity, arising from the anomalous velocity $\frac{1}{\hbar}\nabla \phi \times \boldsymbol{\Omega} (\vec p)$~\cite{dixiao}, is frequency-independent:
\be
\sigma_{xx} = \sigma_{yy} = \frac{n_0e^2/m}{ i \tilde{\omega}}, \quad \sigma_{xy}^{\rm AH} = {\rm const}. 
\label{eq:constitutive}
\ee
Here $\tilde{\omega} = \omega - i/\tau_{\rm tr}$,  
where $\tau_{\rm tr}$ is a transport scattering time that arises, e.g., from impurity scattering. 
We consider collective modes in the regime $\omega \tau_{\rm tr} \gg 1$, where the $1/\tau_{\rm tr}$ contribution to $\sigma_{xx}$ can be neglected.

The anomalous Hall conductivity is easily connected with quantities in the main text via $\sigma_{xy}^{\rm AH} = \mathcal{F} e^2/h$. Here $\mathcal{F} = \sum_i \int d^2 \vec p\, \Omega_i (\vec p) f_{i}^0(\vec p) /  (2\pi \hbar)^2 $ is the dimensionless Berry flux, with $ f_{i}^0(\vec p)$ the equilibrium band occupancy, and $\boldsymbol{\Omega} (\vec p)$ the Berry curvature.Throughout this work we only consider terms up to linear order in $\nabla \phi$.

In this formulation, chiral Berry plasmons (CBPs) at system edges or in disks arise in exactly the same way as shown in the main text, with 
$\bar{\vec v}(\vec r, t)$ and $\mathcal{F}$ traded for $\vec j(\vec r, t)$ and $\sigma_{xy}^{\rm AH}$. 
That is, Eqs.~(\ref{eq:currentcon}) and ~(\ref{eq:constitutive}) of this supplement, with $1/\tau_{\rm tr}\rightarrow 0$, are equivalent to Eqs.~(1) and (2) of the main text;
following the same steps of derivation yields 
Eqs.~(4-11) of the main text with the replacement $\mathcal{F} = h\sigma_{xy}^{\rm AH}/e^2$. 

We note parenthetically that Eq.~(\ref{eq:constitutive}) is different from 
what would be obtained in the analogous case for magnetoplasmons in a system subjected to an applied magnetic field. 
While both cases feature a nonzero Hall conductivity $\sigma_{xy}$, in the case of magnetoplasmons
the applied magnetic field induces a Lorentz force that affects electronic motion by modifying the force balance. 
This leads to different forms of $\sigma_{xx}$ and $\sigma_{xy}$ from those in Eq.~(\ref{eq:constitutive}); in particular, $\sigma_{xy}$ in the presence of a finite magnetic field exhibits an $\omega$ dependence~\cite{fetterB}, in contrast to Eq.~(\ref{eq:constitutive}). This yields qualitatively different behavior of collective modes. 
For example, bulk magnetoplasmons are gapped, whereas bulk plasmons in the presence of a Berry flux giving the same dc-Hall conductivity are gapless (see discussion in main text).

\subsection{CBP dipole mode in a disk}

Here we present a more complete analysis of the CBP dipole modes in a disk, which dominate the optical absorption in metallic disks. These dipole modes can be conveniently described through the center of mass (COM) motion, $\{\vec x (t) \}$, wherein all internal forces cancel (viz.~Newton's third law). Here $\{ \cdot \}$ denotes the COM average, with $\{ \vec x (t)\} = \int d^2 \vec x\, n(\vec x, t) \vec x$, and $\{ \vec p (t) \} = \int d^2 \vec x\, \bar{\vec p}(\vec{x},t)$. The COM equations of motion can be obtained from Eq.~(1) of the main text via integration by parts, along with the condition that the velocity vanishes when $|\vec x| \to \infty$.
This yields 
\be
\partial_t \{ \vec x\}  = \left\{ \frac{\partial \epsilon}{\partial \vec p}\right\} + \frac{1}{\hbar} \{ e\nabla \phi \times \bm \Omega(\vec p) \}, \quad \partial_t \{ \vec p \} = \{e\nabla \phi\},
\label{eq:diskeom-SOI}
\ee
where $\vec E$ is a self-generated electric field associated with the plasmon motion.

To capture the
dipolar mode, we use the harmonic potential $-e\phi(\vec x) = \frac{m}{2} \omega_0^2 |\vec x|^2$ for the electrons 
(see e.g., Ref. \cite{avouris}) where $\omega_0$ is the bare plasmon frequency in a disk (with diameter $d$) in the {\it absence} of Berry curvature. 

In analyzing the COM equations of motion, we first note that $ e \nabla \phi (\vec x)$ depends only on $\vec x$, and $\Omega (\vec p)$ only on $\vec{p}$.  
Keeping terms to linear order in $\delta n$, we find the 
anomalous velocity contribution $\{ \vec v_{a}\}$ for the COM as 
\begin{align}
 \{ e \nabla \phi \times \bm \Omega\}  &= \sum_{\vec x} \Big[e \nabla \phi(\vec x) \times 
\sum_i \int \frac{d^2 \vec p}{(2\pi \hbar)^2} \bm \Omega(\vec p) f_i(\vec x, \vec p , t) \Big] \nonumber \\ 
&=  \!\Big[\sum_{\vec x} e \nabla \phi (\vec x) \mathcal{F} (\vec x, t) \Big] \times \hat{\vec z} \approx \zeta \{\vec x (t) \} \times \hat{\vec{z}},
 \label{eq:avelaverage}
\end{align}
where the Berry flux is given by $\mathcal{F}(\vec x,t ) \equiv \sum_i \int d^2 \vec p\, \Omega_i (\vec p) f_{i}(\vec x, \vec p , t)/ (2\pi \hbar)^2$, the constant $\zeta$ can be obtained as detailed below, and $i$ is a band index. 
Here we will concentrate on the motion of bulk electrons in a given band, $n$.  We therefore take 
$f_{i<n} = 1$ everywhere inside the disk for bands below $n$ (but vanishing outside the disk). 
For the purpose of estimating parameters, we make an assumption of local equilibrium and set  
$f_n (\vec{x}, \vec p, t) =  [e^{\beta(\epsilon_\vec p^n - \mu(\vec x,t))}+1]^{-1}$, where 
 $\mu(\vec x, t)$ is a space and time varying chemical potential. Adopting a simple model of a rigidly moving disk of charge with constant density $n_0$, and density equal to zero outside, gives $\mathcal{F} = \mathcal{F} n(\vec x, t) / n_0$ and $\zeta = -\frac{\mathcal{F}}{n_0} m \omega_0^2$. 
Here $\mathcal{F}$ is obtained with a fixed uniform chemical potential $\mu$. 

Using $\{ e\nabla \phi \} = - m \omega_0^2 \{\vec x (t) \}$, Eq.~(\ref{eq:avelaverage}), and substituting into Eq.~(\ref{eq:diskeom-SOI}), 
we obtain the equations of motion for the dipole mode:  
$(\partial_t^2 + {A}_{ij}) \{x_j\} = 0$, where 
\be
\mathbf{{A}} =  \left( \begin{array}{cc} \omega_0^2  &  \omega_{a} \partial_t   \\ -\omega_{a}  \partial_t  & \omega_0^2 \end{array}\right), \quad \omega_{a} = \frac{\mathcal{F} \omega_0^2 m}{n_0 \hbar}.
\label{eq:eomdisk2}
\ee
Here we have used $ \big\{ \frac{\partial \epsilon}{\partial \vec p} \big\} = \frac{1}{m}\{ \vec p \} + \mathcal{O}(\delta n^2)$ and kept only terms up to linear order in $\delta n$. 

Writing $\{ \vec x \}= \vec x_0 e^{i\omega t}$, we obtain a secular equation  $\mathbf{M} \vec x_0 =0$ where $\mathbf{M} = -\omega^2 + \mathbf{A}$, with $\partial_t$ replaced by $i \omega$ within $\mathbf{A}$. Plasmons are given by the zero modes, ${\rm det}(\mathbf{M})=0$, yielding the split dispersion relation 
\be
\omega_\pm= \sqrt{\omega_0^2 + \frac{\omega_{a}^2}{4}} \pm \frac{\omega_{a} }{2}, 
\label{eq:dispersion-soi}
\ee
where $\omega_\pm, \, \omega_0 >0$. (In the following we will use only positive frequencies; a similar analysis yields the same results for negative $\omega_\pm, \omega_0$ as well). 
These plasmon modes are {\it chiral} as they correspond to a rotating COM 
displacement and momentum 
\be
\{ \vec x (t) \}_\pm = \frac{|\vec x_0|}{\sqrt{2}}\left( \begin{array}{cc} 1 \\ \pm i \end{array} \right)e^{i \omega_\pm t},\, \{ \vec p (t) \}_\pm = \frac{i\omega_0^2 m}{\omega_\pm} \{ \vec x (t) \},\!\!\!
\label{eq:rotation}
\ee 
where $\{ \vec p (t) \}_\pm$ and $\{ \vec x (t) \}_\pm $ are offset by a phase of $\pi/2$. As a result, a rotating $\{ \vec x (t) \}_\pm $ gives rise to a circulating momentum/current density (see Fig.~3a of the main text). The clockwise and anticlockwise ($\omega_+$ and $\omega_-$) motions of chiral plasmons sketched in Fig.~3a are oriented for a positive Berry flux, $\mathcal{F}$, pointing in the $\hat{\vec z}$ direction. This direction sets the orientation of $\omega_{a}$. For opposite sign of $\mathcal{F}$ the orientations are switched. The distinct frequencies $\omega_\pm$ arise from the anomalous velocity $\{\vec v_{a}\}$ adding/subtracting propagation speed from the modes at $\mathcal{F} =0$. 

The CBP dipole mode can manifest in distinct split peaks for optical absorption. This splitting can be analyzed by writing the current density as $\vec j = e n_0  \partial_ t \{ \vec x \}$ by inverting $\mathbf{M}$ above, and relating current to electric field via $\vec j = \vec{g} \vec E$, where $\vec E$ is probing field, and $\vec g$ is the conductivity tensor. Optical absorption is characterized by the real part of the longitudinal conductivity, $\vec g_{\rm xx}$, as  
\be
{\rm Re} [g_{\rm xx}(\omega)] = \frac{1}{2}\sum_\pm\frac{\mathcal{D} \Gamma \omega^2}{(\omega^2 \pm \omega\omega_{a} - \omega_{0}^2)^2 + \Gamma^2\omega^2}, 
\label{eq:sigma}
\ee
where $\mathcal{D} =  \frac{n_0e^2}{m} $ is the Drude weight, and $\Gamma$ is the transport relaxation rate, included phenomenologically via $\partial_t^2 \to \partial_t^2 + \Gamma \partial_t$. This yields the split peak optical absorption for the disk geometry shown in Fig.~3b of the main text. 

\subsection{Optical Selection rules for gapped Dirac materials}

Here we detail how circularly polarized light can yield valley selectivity in gapped Dirac materials (GDMs) where inversion symmetry has been broken. 
This selectivity arises due to the pseudo-spinor nature of the wavefunctions on the A/B sublattices.
The selection rules were derived in Ref.~\cite{yao08} by analyzing the orbital magnetic moments of electronic wavepackets in the $K$ and $K'$ valleys. 
For the reader's convenience, here we present an alternative calculation based on Fermi's golden rule. 

The low energy Hamiltonian for gapped Dirac materials can be described as $\mathcal{H} = H_K + H_{K'}$, where 
\be
\quad H_K = v \bm{\sigma}_+ \cdot \vec p_K + \Delta \sigma_z, \quad H_{K'} = v \bm{\sigma}_- \cdot \vec p_{K'} + \Delta \sigma_z, 
\label{eq:hamiltonian}
\ee 
where $H_K, H_{K'}$ describe electrons close to the $K$ and $K'$ points, $v$ is the Fermi velocity, and $\bm{\sigma}_\pm = \sigma_x \hat{\vec x} \pm \sigma_y \hat{\vec y}$, with $\sigma_{x,y,z}$ the Pauli matrices. Here $\vec p_{K,K'}$ describe momenta taken relative to the $K$ and the $K'$ points, respectively, and $2\Delta$ is the gap size. In the following we shall drop the explicit $K$ and $K'$ labels on $\vec p_{K, K'}$ for brevity. 

A variety of systems obey Eq.~(\ref{eq:hamiltonian}), including van der Waals heterostructures where A/B sublattice symmetry has been broken as in G/hBN heterostructures. 
In such systems the gap size $2\Delta$ corresponds to the asymmetry of the potential on the A/B sublattices. 
Below we will focus on this case for concreteness. However, the underlying physics is general and applies to a broad range of GDMs, such as dual-gated bilayer graphene and transition metal dichalcogenides, where inversion symmetry has been broken. 

The eigenfunctions of Eq.~(\ref{eq:hamiltonian}) can be expressed as pseudo-spinors
\be
| + \ra_{K(K')} = \left( \begin{array}{cc} {\rm cos} \frac{\theta}{2} e^{- i \tau \phi} \\ {\rm sin} \frac{\theta}{2} \end{array}\right), \quad | - \ra_{K(K')} = \left( \begin{array}{cc} {\rm sin} \frac{\theta}{2}e^{- i \tau \phi} \\ -{\rm cos} \frac{\theta}{2} \end{array}\right), 
\label{eq:eigen}
\ee
where $|\pm \ra$ denote states in the conduction (valence) band, $\tau = 1$ for valley $K$ and $\tau = -1$ for valley $K'$, ${\rm tan}\ \theta  = v|\vec p| /\Delta$, and ${\rm tan}\ \phi = p_y/p_x$. Here the energy eigenvalues are $\epsilon_{\vec p}^\pm = \pm \sqrt{v^2 |\vec p|^2 + \Delta^2}$. 

We proceed by noting that when light with frequency $\hbar \omega \geq 2\Delta$ is incident on the GDM, electrons in the valence band can be excited into the conduction band. 
The light-matter coupling is captured 
by writing $\vec p \to \vec p - e\vec A/c$, where the vector potential $\vec{A}$ is related to the incident light electric field $\vec E$ via $\vec A =  \frac{ic}{\omega} \vec E$. Here $\vec{E} = \vec{E}_0 e^{i\omega t}$ for light with frequency $\omega$.
The rate of electron-hole pair creation (absorption of photons) can be calculated via Fermi's golden rule
\be
W_{K(K')} = \frac{2\pi }{\hbar} \sum_{\vec p} |M_{\vec{p}}^{K (K')} |^2 \delta (\epsilon^{+}_{\vec p} - \hbar \omega/2),
\label{eq:rate-1}
\ee
with the matrix elements 
\begin{eqnarray}
\nonumber M_\vec k^{K}  &=& \frac{iev}{\omega} \la + | E_x \sigma_x\ +\ E_y \sigma_y | - \ra_{K}\\
M_\vec k^{K'}\!  &=& \frac{iev}{\omega} \la + | E_x \sigma_x\ -\ E_y \sigma_y | - \ra_{K'}.
\label{eq:MatrixElements}
\end{eqnarray}
We note that the different signs in front of $E_y$ in $ M_\vec k^{K}$ and $M_{\vec{k}}^{K'}$ arise from the different way $\vec A$ couples to pseudospin in $H_K$ and $H_{K'}$ (see Eq.~\ref{eq:hamiltonian}). 

With the help of the identities 
\bea
\la + | \sigma_x | - \ra_{K(K')} &=& {\rm sin}^2 \tfrac{\theta}{2} e^{- i\tau\phi} - {\rm cos}^2\tfrac{\theta}{2} e^{ i \tau \phi}\\
\nonumber\la + | \sigma_y | - \ra_{K(K')} &=& i\big({\rm sin}^2 \tfrac{\theta}{2} e^{- i\tau \phi} + {\rm cos}^2\tfrac{\theta}{2} e^{i \tau \phi}\big),
\eea
and writing $\vec E_0 = |\vec E_0|\, ( \hat{\vec x} + i \eta \hat{\vec y}) /\sqrt{2}$ for left-handed (LH, $\eta = 1$) and right-handed (RH, $\eta = -1$) circularly polarized light, we obtain
\be
W_K^{\eta} = W_0 \left[ \frac{2\Delta}{\hbar \omega} + \eta \right]^2, \ \ W_{K'}^{\eta} = W_0 \left[ \frac{2\Delta}{\hbar \omega} - \eta\right]^2.
\label{eq:rates-2}
\ee
Here $W_0 =N_s e^2 |\vec{E}_0|^2/(16 \hbar^2 \omega) $, where $N_s$ is the spin degeneracy.

Equation (\ref{eq:rates-2}) shows that electron-hole transitions in the valleys $K/K'$ can be selectively excited using LH/RH circularly polarized light (see Fig.~3c inset of the main text). Indeed for $\hbar \omega = 2\Delta$, perfect selection of $K$ or $K'$ electron-hole transitions can be achieved, in agreement with Ref.~\cite{yao08}. We emphasize that this selectivity comes from the {\it orbital} physics of light-matter coupling; it does not require or involve spin-orbit coupling and can even arise in materials with negligible spin orbit-coupling as modeled by Eq.~(\ref{eq:hamiltonian}).

\subsection{Optically pumped ``On-Demand" CBP dipole mode in gapped Dirac materials}

Here we consider how CBPs might arise in non-magnetic gapped Dirac materials such as MoS$_2$, or gapped G/h-BN. To analyze this, we focus on nominally time reversally invariant gapped Dirac materials.
We model the valley dependent $\mathcal{F}$ as~\cite{lensky}:  
$\mathcal{F}_{K,K'}=  \tau_z N_s \frac{\tilde{n}^{1/2}}{ 2({n_{K,K'} + \tilde{n}})^{1/2}}\, {\rm sgn\, \Delta}$, where $n_{K,K'}$ are the valley carrier densities, $\tau_z = \mp$ for the $K,K'$ valleys, $N_s = 2$ is the spin degeneracy, and $\tilde{n} = \Delta^2/{4\pi v_F^2 \hbar^2}$ gives a characteristic density scale. The bandgap is $2|\Delta|$. When $n_K= n_{K'}$, the total flux $\mathcal{F} = \mathcal{F}_K + \mathcal{F}_K'$ vanishes. 

However, a non-vanishing net Berry flux $\mathcal{F}_{\rm pe}$ for the photoexcited system is 
 achieved when $n_K\neq n_{K'}$ (Fig.~3c inset of the main text).
Setting $n_{K'} =0$, giving $\mathcal{F}_{K'} = +1$, 
the valley polarized electron ($n_{\rm el}$) and hole ($n_{\rm h}$) populations concentrated at the $K$ valley conduction and valence band extrema yield
\begin{align}
\mathcal{F}_{\rm pe} &\approx 1- \frac{1}{2} \Big[\frac{\tilde{n}^{1/2}}{\sqrt{\tilde{n} + n_{\rm h}}} +  \frac{\tilde{n}^{1/2}}{\sqrt{\tilde{n} + n_{\rm el}}} \Big] , \nonumber \\ \tilde{n} &= 1.8\times 10^{13} \frac{(\Delta [{\rm eV}])^2 }{(v_F [{\rm cm}\, {s}^{-1}]/10^8)^2}  {\rm cm}^{-2}. 
\label{eq:fluxforTMD}
\end{align}
Here we have used that 
Berry curvature (including its sign) is the same for electrons and holes, following from particle-hole symmetry of the Dirac Hamiltonian, and used the convention $n_{\rm el,h} >0$. 
In deriving Eq.~(\ref{eq:fluxforTMD}), we used the spin degeneracy $N_s = 2$ for the $K'$ valley, and noted that only a single spin species in the $K$ valley is excited by the circularly polarized light.

For demonstration we  
examine CBPs in the disk geometry as above. We analyze the coupled motion of photoexcited electrons and holes in a single valley via COM 
coordinates ($\{\vec x_{\rm el}\}$) and ($\{\vec x_{\rm h}\}$), respectively, giving: 
\be
\Big[\partial_t^2 + 
 \left( \begin{array}{cc} \mathbf{A}_{\rm el}  & - \mathbf{A}_{\rm el} \\ -\mathbf{A}_{\rm h}   & \mathbf{A}_{\rm h} \end{array}\right) \Big]\left( \begin{array}{cc} \{ \vec x_{\rm el} \} \\ \{ \vec{x}_{\rm h} \} \end{array} \right) = 0,
\label{eq:eh}
\ee
where $\mathbf{A}_{\rm el}$  and $\mathbf{A}_{\rm h}$ are defined as in Eq.~(7) of the main text, with the density $n_0$ and plasmon mass $m$ replaced by the appropriate values for electrons or holes.
In writing these equations, we have used that the restoring force arising from the mutual attraction of photoexcited electron and hole subsystems is $\{e \nabla \phi\}_{\rm el(h)} \approx  \mp \alpha (\{\vec x_{\rm el}\} - \{\vec x_{\rm h}\})$, where $\{ \cdot \}_{\rm el,h}$ denote the COM averages for electron and hole distributions and the upper (lower) sign is for electrons (holes). Here $\alpha$ characterizes the strength of the electron-hole interaction. For brevity, in the following analysis we set $\mathbf{A}_{\rm el} = \mathbf{A}_{\rm h}$,
giving $\alpha = m \omega_0^2$, where $\omega_0$ is the plasmon frequency associated with a unipolar system with carrier density $n=n_{\rm el} = n_{\rm h}$. 
 
Zero modes of Eq.~(\ref{eq:eh}) 
with $\partial^2_t \rightarrow -\omega^2$ yield CBPs with 
$\{\vec x_{\rm el}\} = - \{\vec x_{\rm h} \}$, giving:
\be
\omega_\pm^{\rm el-h} = \sqrt{\omega_{a}^2 + 2\omega_0^2} \pm \omega_{a},  
\label{eq:ehplasmons}
\ee
where $\omega_{a}$ is given by Eq.~(7) of the main text with $\mathcal{F}$ replaced by $\mathcal{F}_{\rm pe}$. Importantly, $\omega_\pm^{\rm el-h}$ chiral (electron-hole) plasmons feature 
{\it co-rotating} $\{\vec x_{\rm el}\}$ and $\{\vec x_{\rm h}\}$. This dipole like rotation follows from the equal sign of $\mathcal{F}$ for electrons and holes. Modes with $\{\vec x_{\rm el}\} = \{\vec x_{\rm h} \}$ are also eigenmodes of Eq.~(\ref{eq:eh}). However, they have frequency, $\omega =0$.


\begin{thebibliography}{99}

\bibitem{nagaosa} Nagaosa N, Sinova J, Onoda S, MacDonald AH, Ong NP (2010) Anomalous Hall effect.
{\it Rev Mod Phys} {82}, 1539-1592.


\bibitem{haldane} Haldane FDM (1988), Model for a Quantum Hall Effect without Landau Levels: Condensed-Matter Realization of the ``Parity Anomaly,'' {\it Phys. Rev. Lett.} {61}, 2015.
\bibitem{yu2010} Yu R., et al (2010) Quantized anomalous Hall effect in magnetic topological insulators {\it Science} {329}, 61.
\bibitem{nagaosa2011} Nomura K, Nagaosa N (2011) Surface-quantized anomalous Hall current and the magnetoelectric effect in magnetically disordered topological insulators {\it Phys. Rev. Lett.} {106}, 166802.

\bibitem{qahe} Chang C-Z, et al. (2013) Experimental Observation of the Quantum Anomalous Hall Effect in a Magnetic Topological Insulator, {\it Science} {34}, 1629.
\bibitem{wang14} Wang Q-Z, Liu X, Zhang HZ, Samarth N, Zhang S-C, Liu C-X (2014) Quantum Anomalous Hall Effect in Magnetically Doped InAs/GaSb Quantum Wells, {\it Phys. Rev. Lett.} {113}, 147201.

\bibitem{mak2014} Mak KF, McGill KL, Park J, McEuen PL (2014) The valley Hall effect in MoS${}_2$  transistors. 
{\it Science} {344},1489-1492.

\bibitem{dixiao} Xiao D, Meng MC, Niu Q (2010) Berry phase effects on electronic properties, {\it Rev. Mod. Phys.} {82}, 1959-2007.


\bibitem{hasan} Hasan MZ, Kane CL (2010) Colloquium: Topological insulators. {\it Rev. Mod. Phys.}  {82}, 3045.


\bibitem{joannopolus} Jalas D, et al. (2013) What is - and is not- an optical isolator, {\it Nat. Photonics} {7}, 579.

\bibitem{barnes-plasmons} Barnes WL, Dereux A, Ebbesen TW (2003) Surface plasmon subwavelength optics, {\it Nature} {424}, 828. 
\bibitem{qplasmonics} Tame MS, McEnery KR, Ozdemir SK, Lee J, Maier SA, Kim MS (2013), Quantum Plasmonics, {\it Nat. Phys.} {9}, 329.
\bibitem{polinireview} Grigorenko AN, Polini M, Novoselov KS (2012) Graphene Plasmonics, {\it Nat. Photonics} {6}, 749.


\bibitem{stormer} Allen SJ, St\"ormer HL, Hwang JCM (1983) Dimensional resonance of the two-dimensional electron gas in selectively
doped GaAs/A$_1$GaAs heterostructures, {\it Phys. Rev. B}, {28} 4875.  


\bibitem{magnetoplasmonGAP} See e.g. Theis TN (1980) Plasmons in inversion layers, {\it Surf. Sci.} {98}, 515. Heitman D (1986), Two-dimensional plasmons in homogeneous and laterally microstructured space charge layers, {\it Surf. Sci.}, {170}, 332.

\bibitem{fetterexp} Mast DB, Dahm AJ, Fetter AL (1985) Observation of Bulk and Edge Magnetoplasmons in a Two-Dimensional Electron Fluid, {\it Phys. Rev. Lett.} {54}, 1706.
\bibitem{evaandrei} Glattli DC, Andrei EY, Deville G, Poitrenaud J, Williams FIB (1985) Dynamical Hall Effect in a Two-Dimensional Classical Plasma, {\it Phys. Rev. Lett.}, {54}, 1710.

\bibitem{kohn1961} See e.g. Kohn W (1961), Cyclotron Resonance and de Haas-van Alphen Oscillations of an Interacting Electron Gas, {\it Phys. Rev.} {123} 1242; Brey L, Johnson NF, Halperin BI (1989) Optical and magneto-optical absorption in parabolic quantum wells, {\it Phys. Rev. B } {40}, 10647-10649; Maksym PA, Chakraborty, T (1990) Quantum dots in a magnetic field: Role of electron-electron interactions, {\it Phys. Rev. Lett.} {65}, 108.

\bibitem{fetterA} Fetter AL (1985) Edge magnetoplasmons in a bounded two-dimensional electron fluid, {\it Phys. Rev. B} {32} 7676.


\bibitem{pkbook} Lifshitz EM, Pitaevskii LP, (1981) Physical Kinetics, {\it Butterworth-Heinemann}.
\bibitem{guliani} Giuliani GF, Vignale G (2005), Quantum Theory of the Electron Liquid, {\it Cambridge} 

\bibitem{appendix} See Supplementary Online Information (SOI) for discussions of: an alternative formulation of chiral Berry plasmons based on current densities and the conductivity tensor, chiral Berry plasmon dipole modes in a disk, optical selection rules for gapped Dirac materials with inversion symmetry breaking, and photo-induced chiral Berry plasmons.

\bibitem{donheeham} Yoon H, et al. (2014) Measurement of collective dynamical mass of Dirac fermions in graphene. {\it Nat. Nano.} {9}, 594-599.

\bibitem{fetterB} Fetter AL (1986) Edge magnetoplasmons in a two-dimensional electron fluid confined to a half-plane, {\it Phys. Rev. B} {33}, 3717.

\bibitem{Mikhailov} Volkov VA and Mikhailov SA (1988) Edge magnetoplasmons: low frequency weakly damped excitations in inhomogeneous two-dimensional electron systems, {\it Zh. Eksp. Teor. Fiz.} {94}, 217.

\bibitem{wang11} Wang W, Apell P, Kinaret J (2011) Edge plasmons in graphene nanostructures, {\it Phys. Rev. B}  {84}, 085423.


\bibitem{cooper} Price HM, Cooper NR (2013) Effects of Berry Curvature on the Collective Modes of Ultracold Gases, {\it Phys. Rev. Lett.} {111}, 220407.

\bibitem{avouris} Yan H, Li Z, Li X, Zhu W, Avouris P, Xia F (2012) Infrared Spectroscopy of Tunable Dirac Terahertz Magneto-Plasmons in Graphene, {\it Nano Lett.} {12}, 3766.


\bibitem{culcer} Culcer D, MacDonald AH, Niu Q (2003), Anomalous Hall effect in paramagnetic two-dimensional systems, {\it Phys. Rev. B} {68}, 045327.

\bibitem{yao08} Yao W, Xiao D, Niu Q (2008) Valley-dependent optoelectronics from inversion symmetry breaking, {\it Phys. Rev. B} 77, 235406.

\bibitem{lensky} Lensky YD, Song JCW, Samutpraphoot P, Levitov LS (2015) Topological Valley Currents in Gapped Dirac Materials, {\it Phys. Rev. Lett.} {114}, 256601.



\bibitem{jones} Jones AM, et al. (2013) Optical generation of excitonic valley coherence in monolayer WSe2, {\it Nat. Nano.} {8},  634-638.

\bibitem{heinz_mote2} Ruppert C, Aslan OB, Heinz TF (2014) Optical Properties and Band Gap of Single-and Few-Layer MoTe$_2$ Crystals, {\it Nano Lett.} {14}, 6231-6236.


\bibitem{hunt13} Hunt B, et al. (2013) Massive Dirac Fermions and Hofstadter Butterfly in a van der Waals Heterostructure.
{\it Science} {40}, 1427.

\bibitem{woods14} Woods CR, et al. (2014) Commensurate-incommensurate transition in graphene on hexagonal boron nitride, {\it Nat. Phys.} {10}, 451-456.


\bibitem{gorbachev14} Gorbachev RV, et al. (2014), Detecting Topological Currents in Graphene Superlattices, {\it Science} {346}, 448.
\bibitem{bilayergraphene} Zhang Y, Tang T-T, Girit C, Hao Z, Martin MC, Zettl A, Crommie MF, Ron Shen Y, Wang F (2009) Direct observation of a widely tunable bandgap in bilayer graphene, {\it Nature} {459}, 820-823.



\bibitem{wangyao07} Yao W, MacDonald AH, Niu Q (2007) Optical Control of Topological Quantum Transport in Semiconductors, {\it Phys. Rev. Lett.}  {99}, 047401.

\bibitem{kitagawa} Kitagawa T, Oka T, Brataas A, Fu L, Demler E (2011) Transport properties of nonequilibrium systems under the application of light: Photoinduced quantum Hall insulators without Landau levels, {\it Phys. Rev. B} {84}, 235108.
\bibitem{kumar} Kumar A, Nemilentsau A, Fung KH, Hanson G, Fang NX, Low T, Chiral plasmon in gapped Dirac systems, {\it Phys. Rev. B} {\bf 93} 041413 (2016). 

\bibitem{Henriksen} Henriksen E. A., et al. (2010) Interaction-Induced Shift of the Cyclotron Resonance of Graphene Using Infrared Spectroscopy, {\it Phys. Rev. Lett.} {104} 067404.


\bibitem{yao14} Yu H, Liu G-B, Gong P, Xu X, Yao W (2014) Dirac cones and Dirac saddle points of bright excitons in monolayer transition metal dichalcogenides, {\it Nat. Comm.} {5}, 3876.

\end{thebibliography}
\end{document}